\documentclass[prl,showpacs,twocolumn]{revtex4}
\usepackage{epsfig}
\catcode`\"=\active \let"=\" \let\3=\ss
\begin{document}
\bibliographystyle{revtex}
\title{Charge and spin dynamics of interacting Fermions in a
one dimensional harmonic trap}
\author{Lars Kecke}
\affiliation{Physikalisches Institut, Universit\"at Freiburg,
Hermann-Herder-Str.~3, D-79104 Freiburg, Germany}
\author{Wolfgang H\"ausler}
\affiliation{Physikalisches Institut, Universit\"at Freiburg,
Hermann-Herder-Str.~3, D-79104 Freiburg, Germany}
\affiliation{Institut f\"ur Theoretische Physik,
Universit\"at Erlangen, Staudtstr.~7, D-91058 Erlangen,
Germany}
\author{Hermann Grabert}
\affiliation{Physikalisches Institut, Universit\"at Freiburg,
Hermann-Herder-Str.~3, D-79104 Freiburg, Germany}

\begin{abstract}
We study an atomic Fermi gas interacting through repulsive contact
forces in a one dimensional harmonic trap. Bethe-Ansatz solutions 
lead to an inhomogeneous Tomonaga-Luttinger model for the low energy
excitations. The equations of motion for charge and spin density
waves are analyzed both near the trap center and near the trap edges.
While the center shows conventional spin-charge separation the edges
cause a giant increase of the separation between these modes.
\end{abstract}
\maketitle
Recent advances in cooling technology have allowed to reach the degenerate
regime of Fermionic quantum gases. Different hyperfine states effectively
correspond to spin polarizations of spin-$\frac{1}{2}$ particles
of respective densities $n_{\uparrow}$ and $n_{\downarrow}$ \cite{demarco}.
Furthermore, the spatial dimensionality of the gases can be reduced using
either a hollow beam setup \cite{bongs} or arrays of microtraps
\cite{reichel} so that one dimensional (1D) gases can be studied, where
interaction effects are known to be most pronounced. As a consequence of
pure $s$-wave scattering in 3D \cite{olshanii} the interaction is short
ranged \cite{morigi} and acts only between particles of opposite spins.
Also, the strength of the forces can, in principle, be varied over wide
ranges \cite{loftus} by tuning the Feshbach resonance \cite{feshbach}. An
optical lattice along the trap can further enhance correlations by
reducing the bandwidth and thus the kinetic energy
\cite{zwerger,esslinger}.

As opposed to higher dimensions, total charge
($\rho=n_{\uparrow}+n_{\downarrow}$) and relative spin
($\sigma=n_{\uparrow}-n_{\downarrow}$) density waves comprise the only low
energy excitations in 1D; no Fermionic quasiparticles can be excited in
this regime. The Tomonaga-Luttinger liquid (TLL) theory
\cite{tomolut,haldane} describes the properties of homogeneous 1D systems
entailing that charge and spin modes move at different velocities when
interactions are present. This spin-charge separation is considered as a
hallmark of TLL behavior. A parabolic trap potential, however,
causes the particle density to vary along the trap. To treat such an 
inhomogeneous gas cloud, new Boson representations of the Fermi operators 
have been introduced \cite{wonneberger}. Here, we follow the other route 
put forward recently by Recati et al.\ \cite{zwerger} and
consider an inhomogeneous TLL \cite{schulz} with $x$-dependent parameters, assuming
a trap potential which is slowly varying on the scale of the Fermi wavelength.

In the bulk of the gas cloud, away from the trap edges,
interactions are sufficiently weak to justify perturbative
estimates to the (local) TLL model parameters. In this regime,
the picture of charge and spin waves moving at different velocities
is recovered \cite{jagla,zwerger}. Near the trap edges,
however, the gas density decreases and interactions become
very strong. In this regime we employ the Bethe-Ansatz solution for
spin-$\frac{1}{2}$ particles with contact forces
\cite{yang,liebwu,takahashi} to compute the TLL parameters
exactly \cite{shiba,coll}. To leading order in the inverse
interaction strength we can access the dynamics analytically.
Charge modes are found to propagate in a similar way as in
the bulk of the gas cloud. At the edges they are reflected and
thus keep oscillating in the trap until damping processes
become significant. Remarkably, spin density waves show an
exponential slowing down of their velocity and, ultimately,
accumulation at an edge without reflection. Thus, Fermionic
1D quantum gases in shallow confinements should exhibit a giant
increase of spin-charge separation, much more pronounced than
electrons in quantum wires \cite{voit}. The effect of spin
accumulation should be detectable: one way would be to observe
(e.g.\ by fluorescence measurements) the time evolution of an initial
spin-up density peak containing charge and spin modes of equal
amounts, $\rho_\uparrow=(\rho+\sigma)/2$.

Without parabolic confinement the quantum gas is described by
the Hamiltonian $(\hbar=1)$
\begin{equation}
H_{{\rm hom}}=\frac{1}{2m}\left[\sum_{i=1}^N-\frac{\partial^2}
{\partial x_i^2}+2c\sum_{i<j}\delta(x_i-x_j)\right]\,.
\label{Hhom}\end{equation}
Effectively, by virtue of the Pauli principle, only Fermions of
opposite spins are interacting. As long as the 3D
scattering length $a$ is much smaller than the transversal
width $d$ of the gas cloud, the interaction strength $c=2a/d^2$
\cite{olshanii}.

\begin{figure}
\centerline{\epsfig{file=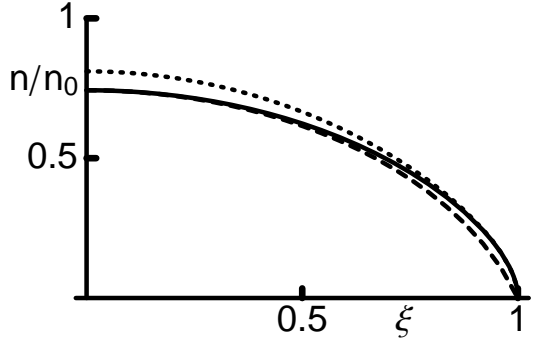,width=8cm}}
\caption{LDA density profile of interacting Fermions in a trap
of length $2R$ for dimensionless interaction strength $u=0.3$,
using the full Bethe-Ansatz equations numerically (solid line),
Eq.~(\ref{ncent}) (dashed line),
and Eq.~(\ref{nedge}) (dotted line); $n_0=\frac{2}{\pi}
m\omega_{{\rm T}} R$ is the density of a non-interacting Fermi gas
at the same $\mu$.}
\label{dichte}
\end{figure}

For given particle density $n$, the ground state energy density
$\epsilon_{{\rm hom}}(n)$ is obtained exactly from the Hamiltonian
(\ref{Hhom}) via the Bethe-Ansatz \cite{liebwu}.
In presence of the longitudinal trap potential $V(x)$, we determine
the inhomogeneous profile of the particle density $n(x)$ in the spirit
of the local density approximation (LDA) \cite{zwerger} by minimizing
\begin{equation}
E[n]=\int dx\left[\epsilon_{{\rm hom}}(n(x))+
V(x)n(x)-\mu n(x)\right]\;.
\label{lda}\end{equation}
This is adequate for slow variations of $V$
compared to the Fermi wavelength and for large particle numbers,
$N\gg 1$, where we can ignore Friedel oscillations occurring on the scale
of the Fermi wavelength $\pi/2k_{{\rm F}}$ \cite{wonneberger}.
The chemical potential $\mu$ ensures that $N=\int dx\;n(x)$ and
determines the length $2R$ of the gas cloud through $V(R)=\mu$,
independent of $V(|x|<R)$ and the form of
$\epsilon_{{\rm hom}}(n)$, provided
$\partial_n\epsilon_{{\rm hom}}(n=0)=0$ \cite{astra}. Assuming now a
harmonic trap potential $V(x)=\frac{m}{2}\omega_{{\rm T}}^2x^2$ of
frequency $\omega_{{\rm T}}$, one finds from Eq.~(\ref{lda})
\begin{equation}
\partial_n\epsilon_{{\rm hom}}(n(x))+
\frac{m}{2}\omega_{{\rm T}}^2x^2-\mu=0\;.\label{ldan}\end{equation}
The resulting density profile $n(x)$ is depicted in Fig.~\ref{dichte}
\cite{olshanii01}.

To proceed we now focus on two regions of particular interest.
Near the center of the trap, where the density $n$ is large enough
to validate perturbative expressions in the interaction strength, 
one finds to leading order in $c/n$ \cite{mahan} 
\begin{equation}
\epsilon_{{\rm pert}}(n)=\frac{\pi^2n^3}{24m}
\left(1+\frac{c}{\pi^2n}\right)\;. \label{epert}\end{equation}
Inserting (\ref{epert}) into (\ref{ldan}) gives a modified
Thomas-Fermi profile
\begin{equation}
n_{{\rm pert}}(\xi)=\frac{2m}{\pi}v_{{\rm F}0}
\left(\sqrt{1+u^2-\xi^2}-u\right)\;.
\label{ncent}\end{equation}
Here we have introduced the dimensionless
coordinate $\xi=x/R$ and interaction strength $u=c/m\pi v_{{\rm F}0}$
where $v_{{\rm F}0}=R\omega_{{\rm T}}$ is the Fermi velocity of a
homogeneous, non-interacting system at chemical potential $\mu$.
On the other hand, near the trap edges, $c\gg n$ so that we can
evaluate the Bethe-Ansatz solution to leading order in $n/c$,
yielding \cite{zwerger}
\begin{equation}
\epsilon_{{\rm edge}}(n)=\frac{\pi^2n^3}{6m}
\left(1-\frac{4n}{c}\ln 2\right)\;.
\label{eba}\end{equation}
Inserting (\ref{eba}) into (\ref{ldan}) gives
\begin{equation}
n_{{\rm edge}}(\xi)=\frac{2m}{\pi}\frac{v_{{\rm F}0}}{2}
\left(\sqrt{1-\xi^2} +\frac{8\ln 2}{3\pi^2u}(1-\xi^2)\right)\;.
\label{nedge}\end{equation}
As seen in Fig.~\ref{dichte}, the analytic forms (\ref{ncent})
and (\ref{nedge}) describe the trap
density accurately in their regions of validity.
We stress that $n_{{\rm edge}}(\xi)$ vanishes with infinite slope
at the edges $\xi\to\pm 1$, which is not seen from the
perturbative expression $n_{{\rm pert}}(\xi)$.

The inhomogeneous TLL Hamiltonian is of the form
\begin{equation}\label{htl}
H_{{\rm TLL}}=\frac{1}{2}\sum_{\nu=\rho,\sigma}\int dx
\big[v_{{\rm J}\nu}(x)(\Theta_\nu'(x))^2+v_{{\rm N}\nu}(x)(\Phi_\nu'(x))^2
\big]\;,
\end{equation}
where primes denote spatial derivatives, and $\Theta_\nu$ and
$\Phi_\nu$ are the usual Bosonic phase fields in the charge $(\nu=\rho)$
and spin $(\nu=\sigma)$ sectors, obeying
$[\Phi_\nu(x),\Theta_{\nu'}'(x')]={\rm i}\delta_{\nu\nu'}\delta(x-x')$.
They define density excitations
$\nu=\langle\Phi_\nu'\rangle/\sqrt{\pi}$.
The velocity parameters $v_{{\rm N}\nu}(x)$ follow from exact
compressibility relations \cite{haldane,voit,whlk}
\begin{equation}
v_{{\rm N}\nu}=\frac{2}{\pi}\frac{\partial^2\epsilon_{{\rm hom}}}
{\partial\nu^2}\;,\label{vofe}
\end{equation}
while symmetries fix the values of velocity parameters $v_{{\rm J}\nu}$:
$v_{{\rm J}\rho}=\pi n/2m$ owing to Galilei invariance and
$v_{{\rm J}\sigma}=v_{{\rm N}\sigma}$ as a consequence of SU(2)
invariance in the spin sector. The model (\ref{htl}) can be justified if
$\nu\ll n$ which, evidently, fails too close to the trap edges where
$n\to 0$, see below.

After separating off the time dependence $\sim{\rm e}^{{\rm i}\omega t}$
in the Heisenberg equations of motion, the eigenvalue equations for the
density excitations become
\begin{eqnarray}\label{dyn}
\mbox{}-v_{{\rm N}\nu}v_{{\rm J}\nu}\nu''
-\left(2v'_{{\rm N}\nu}v_{{\rm J}\nu}
+v_{{\rm N}\nu}v'_{{\rm J}\nu}\right)\nu'&&\\
\mbox{}-\left(v'_{{\rm N}\nu}v'_{{\rm J}\nu}
+v''_{{\rm N}\nu}v_{{\rm J}\nu}\right)\nu&=&
\omega^2\nu\;.\nonumber
\end{eqnarray}
As boundary conditions we require that
charge and spin currents vanish at the trap edges, $j_\nu(x=\pm
R)=0$. In view of (\ref{htl}) and the identity
$j_\nu=\dot\Theta_\nu/\sqrt{\pi}$, this boundary condition becomes
$v_{{\rm J}\nu}v_{{\rm N}\nu}'\nu+v_{{\rm J}\nu}v_{{\rm N}\nu}\nu'=0$
at $\xi=\pm 1$ which, together with (\ref{dyn}), governs the 
dynamics of density wave packets. In the homogeneous case
$v'_{{\rm (N/J)}\nu}$ and $v''_{{\rm (N/J)}\nu}$ vanish and (\ref{dyn})
simplifies to a wave equation. In the inhomogeneous case, without
interactions, all four TLL velocity parameters are equal and given by
$v_{{\rm F}}(\xi)=v_{{\rm F}0}\sqrt{1-\xi^2}$ which follows from
Eqs.~(\ref{ldan}) and~(\ref{vofe}) for $\epsilon_{{\rm hom}}(n)=
\pi^2n^3/24m$. Then the solutions of (\ref{dyn}) \cite{minguzzi}
\begin{equation}
\nu_{{\rm nonint}}(x,\omega)=\frac{1}{\sqrt{R^2-x^2}}
\cos\left(\bar\omega\:{\rm arccos}\:\xi\right)\, ,\label{nonint}
\end{equation}
with integer $\bar\omega=\omega/\omega_{{\rm T}}$, constitute a complete
orthonormal set with regard to the measure $\frac{2R}{\pi\sqrt{R^2-x^2}}$
\cite{sturmliouville}.

Near the trap center, we obtain from Eqs.~(\ref{ncent}) and~(\ref{vofe})
to leading order in $u$ 
\begin{eqnarray}\label{vpert1}
v_{{\rm J}\rho}&=&v_{{\rm F}}
=v_{{\rm F}0}\left(\sqrt{1+u^2-\xi^2}-u\right)\\
v_{{\rm N}\rho}&=&v_{{\rm F}}+u v_{{\rm F}0}
=v_{{\rm F}0}\sqrt{1+u^2-\xi^2}\label{vpert2}\\
v_\sigma&=&\sqrt{v_{{\rm F}}(v_{{\rm F}}-uv_{{\rm F}0})}
\approx v_{\sigma 0}\sqrt{1-x^2/R_\sigma^2}\;,
\label{vpert3}\end{eqnarray}
where $v_{\sigma 0}=v_{{\rm F}0}\sqrt{1-3u}$ and
$R_\sigma=R\sqrt{1-3u/2}$. The solutions of (\ref{dyn}) now become
\begin{eqnarray}\label{rhonx}
\rho_n(x)&=&\frac{n/2}{\sqrt{(1+u^2)R^2-x^2}}{\rm C}_n^{\left(h/2\right)}
\left(\frac{x}{R_2}\right)\\
\sigma_n(x)&=&\frac{1}{\sqrt{R_\sigma^2-x^2}}\cos
\left(n\:{\rm arccos}\:\frac{x}{R_\sigma}\right)\;,
\label{sigmanx}\end{eqnarray}
where the ${\rm C}_n^{(\alpha)}$ denote ultraspherical
polynomials \cite{AS}, $R_2=R\sqrt{1-u/2}$, $h=u/(2-u)$, and the
eigenfrequencies are $\omega_{\rho,n}=\omega_{{\rm T}}\sqrt{n(n+h)/(1+h)}$,
$\omega_{\sigma,n}=\omega_{{\rm T}} n\sqrt{(1-3u)(1-3u/2)}$. Thus, in
the charge sector the spectrum is no more
equidistant (see also Refs.~[\onlinecite{zwerger,astra}]),
although the lowest mode ($n=1$) remains unaffected by interactions
in accordance with Kohn's theorem \cite{kohn}. 
Eqs.~(\ref{rhonx}) and~(\ref{sigmanx}) describe how weak
interactions smoothly deform the non-interacting solution
(\ref{nonint}); note that
${\rm C}^{(0)}_n(\xi)=\frac{2}{n}\cos(n\:{\rm arccos}\:\xi)$. The
inequalities $R_2\omega_{\rho,n}>n v_{{\rm F}0}$ and 
$R_\sigma\omega_{\sigma,n}<n v_{{\rm F}0}$ reflect the enhanced and
suppressed dynamics of charges and spins, respectively.

Near the trap edges, in the non-perturbative regime where
$\sqrt{R^2-x^2}\ll uR$ (but still $R-|x|\gg k_{{\rm F}}^{-1}$),
interactions alter the above picture, even qualitatively. Using
Eqs.~(\ref{eba}) and~(\ref{nedge}), and employing the Bethe-Ansatz
result by Coll \cite{coll}, we get from (\ref{vofe}) to leading
order in $u^{-1}\sqrt{1-\xi^2}$
\begin{eqnarray}
v_{{\rm J}\rho}&=&v_{{\rm F}0}[\sqrt{1-\xi^2}/2+
\beta(1-\xi^2)/2]\label{vj}\\
v_{{\rm N}\rho}&=&v_{{\rm F}0}[2\sqrt{1-\xi^2}-
4\beta(1-\xi^2)]\label{vn}\\
v_\sigma&=&\frac{4\pi m v_{{\rm F}}^2}{3c}=
\frac{v_{{\rm F}0}}{3u}(1-\xi^2)
\label{vs}
\end{eqnarray}
where $\beta=8\ln 2/(3\pi^2u)$. Disregarding for the moment
subleading terms $\propto\beta$ in Eqs.~(\ref{vj}) 
and (\ref{vn}), we see that the charge sector exponent
$K_{\rho}=\sqrt{v_{{\rm J}\rho}/v_{{\rm N}\rho}}$ approaches the
value $1/2$ close to the edges, which is consistent with the limiting
value of $K_{\rho}$ in the Hubbard model when the filling goes to
zero \cite{schulz90}. The plasmon velocity
$\sqrt{v_{{\rm N}\rho}v_{{\rm J}\rho}}$, on the other hand, takes
exactly the value of the non-interacting system [cf.\
Eqs.~(\ref{vpert1}) and (\ref{vpert2}) for $u\to 0$]. This amazing
property originates from the fact that the particle density
is reduced by a factor $1/2$ at infinite interactions, compared
to the non-interacting system.
Furthermore, since Eq.~(\ref{dyn}) only
contains products $v_{{\rm N}\nu}v_{{\rm J}\nu}$ and their spatial
derivatives, we conclude that even the whole charge dynamics
coincides exactly with the non-interacting dynamics (\ref{nonint}) 
sufficiently close to the trap edges.

To leading non-trivial order in $\beta$, Eq.~(\ref{dyn}) can be
transformed into a damped Mathieu equation, that is solved by
\begin{equation}
\rho_n(x)=\frac{\exp(\frac{3}{2}\beta\sqrt{1-\xi^2})}{\sqrt{R^2-x^2}}
\left\{{se_{2n}(q,\frac{1}{2}\arcsin\xi)\,,n\,\mbox{odd}}
\atop{ce_{2n}(q,\frac{1}{2}\arcsin\xi)\,,n\,\mbox{even}}\right\}
\label{rhomat}\end{equation}
with the Mathieu functions $se_n$ and $ce_n$. The eigenenergies
$\omega_n$ follow from the characteristic values \cite{AS}
$a_{2n}(q_n)=4\omega_n^2-9\beta^2$ of the Mathieu functions while
$q_n=-\beta(2\bar\omega_n^2-1)$. When $\beta\sqrt{1-\xi^2}\to
0$, the $\rho_n(x)$ turn into the non-interacting solutions
(\ref{nonint}), in agreement with the above observations for
the vicinity of the trap edges.

In the spin sector the corresponding solutions,
\begin{equation}
\sigma(x,\omega)=\frac{R}{R^2-x^2}\left\{\sin\atop\cos\right\}
\left(\frac{3u\omega}{\omega_{{\rm T}}}\:{\rm Atanh}\:\xi\right)\;,
\end{equation}
differ qualitatively from the non-interacting solutions
(\ref{nonint}), as well as from the weakly interacting ones (\ref{sigmanx}).
They are rapidly oscillating near the trap edges which
manifests extremely slow spatial propagation of spin density waves,
as we discuss now.

With the eigenfunctions for charge and spin densities at hand, 
we are in the position to solve for the time evolution of an initial
wave packet $\nu_0(x)$ in the trap center, or near the edge. 
After preparation, the packet splits into left-
and right moving parts, $\nu_+$ and $\nu_-$, and 
the charge and spin constituents separate. Near the edge the time
evolution follows
\begin{eqnarray}
\rho_\pm(x,t)&\approx&\frac{1}{2}\sqrt{\frac{R^2-x'^2}{R^2-x^2}}
\rho_0(x')\nonumber\\
x'&=&R\cos\left({\rm arccos}\:\xi\pm\omega_{{\rm T}} t\right)\label{rhopm}\\
\sigma_\pm(x,t)&=&\frac{1}{2}\frac{R^2-x'^2}{R^2-x^2}\sigma_0(x')\nonumber\\
x'&=&R\tanh\left({\rm Atanh}\:\xi\pm\frac{\omega_{{\rm T}}}{3u}t\right)\;.
\label{sigmapm}\end{eqnarray}
Subleading corrections of order $\beta$ require a summation over
the eigenfunctions (\ref{rhomat}) and can no longer be expressed in
the simple form (\ref{rhopm}).

\begin{figure}
\centerline{\epsfig{file=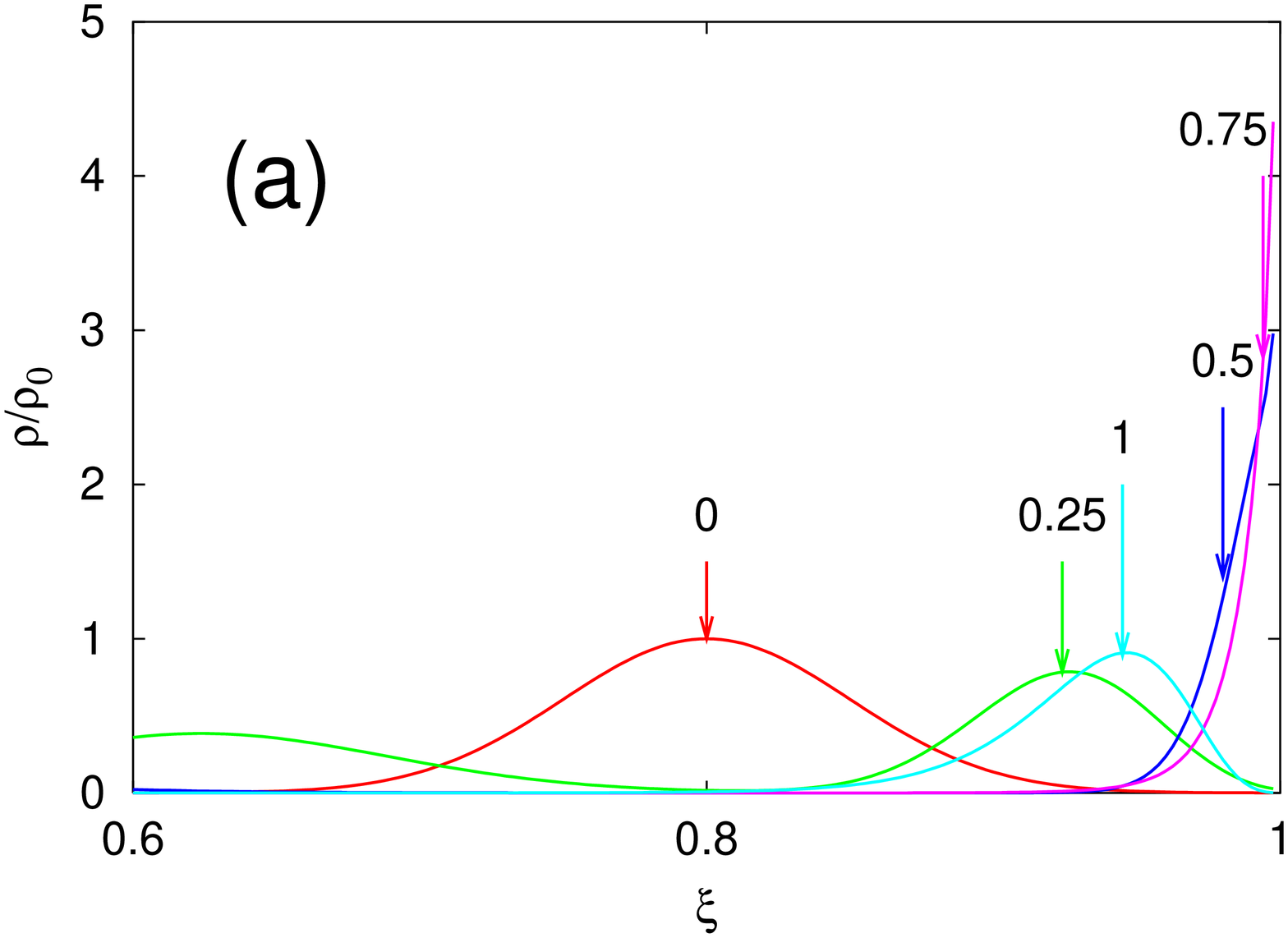,angle=270,width=7cm}}
\centerline{\epsfig{file=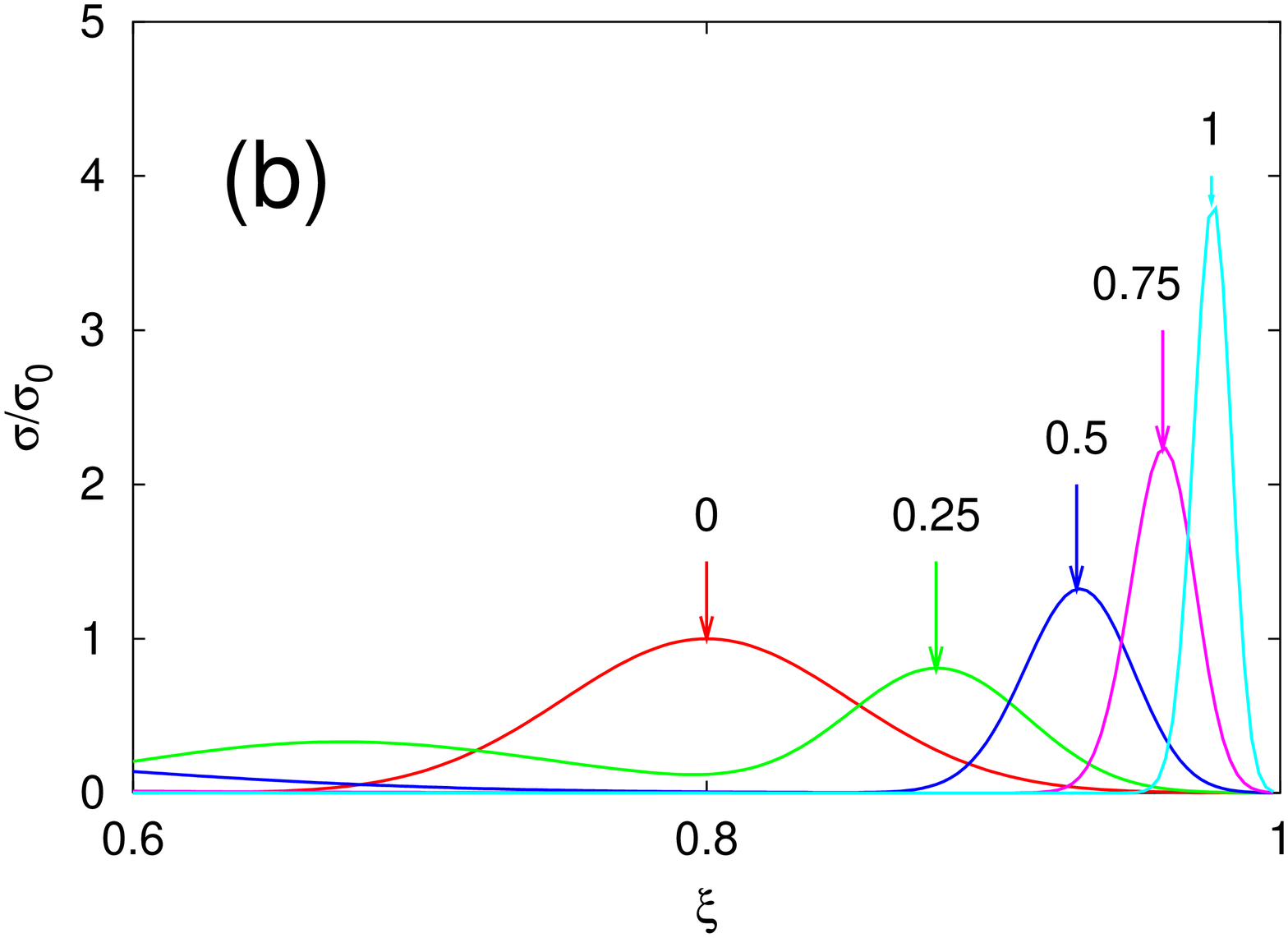,angle=270,width=7cm}}
\caption{(Color online) Propagation of a charge density (a) and a
spin density (b) excitation for dimensionless interaction strength 
$u=0.3$. Time steps are $\Delta t\omega_{{\rm T}}=0.25$ as
indicated. The charge wave is reflected at the edge of the trap,
similar to the non-interacting case, while the
spin wave exponentially slows down without reflection.}
\label{evfig}
\end{figure}

It follows from Eq.~(\ref{rhopm}) that the charge density exhibits
a temporal evolution that roughly resembles that in the
absence of interactions: the right moving part of the initial
wave packet slows down somewhat due to the decreasing Fermi
velocity, and then is reflected at the edge (cf.\ Fig.~\ref{evfig}a).
We thus expect
charge density excitations to keep oscillating in the trap,
before damping mechanisms set in. On the contrary, the right
moving part of the spin density wave is slowed down even exponentially
fast, see Eq.~(\ref{sigmapm}), and is not reflected 
(cf. Fig.~\ref{evfig}b). We mention that for long times 
Eq.~(\ref{sigmapm}) gives $\sigma_\pm(x,t\to\infty)\to A_0
\delta(x\pm R)$ where $A_0=\int dx\:\sigma_0(x)$
is the conserved magnitude of the initial spin wave packet.
This follows from the finite slope of $v_\sigma$
at $\xi=\pm 1$, as opposed to the infinite slope $v'_{\rho{\rm (N/J)}}$ at 
$\xi=\pm 1$ in the charge sector. However, our analysis is only valid
up to times of order $t_c\approx\frac{u}{\omega_{{\rm T}}}
\left[\ln\left(\frac{2mR\omega_{{\rm T}}}
{\pi\sigma_0(\xi_0)}\sqrt{1-\xi_0^2}\right)+3\ln\frac{2}{1+\xi_0}\right]$
since then $\sigma_\pm(x,t>t_c)$ exceeds $n(x)$. This time is large for small initial
displacements $\sigma_0$ so that a dramatic separation of the spin and
charge peaks will have occured well before effects on the length scale of
the Fermi wavelength, that have been disregarded here, become relevant.

In conclusion, we have studied interacting Fermionic atoms of
two spin species in an effectively one dimensional harmonic trap
by exploiting solvability by the Bethe-Ansatz for contact forces.
We have evaluated the particle density profile within the local
density approximation. Near the trap center we confirm the
occurrence of spin-charge separation. Near the edges of the trap
interactions affect the dynamics in an unexpectedly drastic way.
While in the immediate vicinity of an edge charge density waves
move as if they were non-interacting with reflection at the
edge, spin density waves are not reflected and accumulate at the
edge. This establishes an even more pronounced spin-charge
separation than in homogeneous systems. Experimentally one
could confirm our predictions e.g.\ by selectively evaporating
parts of the `up' spin component $\rho_\uparrow$ in a small
region of space and observe the right and left moving spin and
charge density waves separating from one another (since
$2\rho_\uparrow=\rho+\sigma$), both in the bulk and near the
edge of the trap.

We thank R. Egger, A. Komnik and W. Zwerger for valuable comments.


\begin{thebibliography}{99}
\bibitem{sturmliouville} Note that the eigenvalue problem of
Eq.~(\ref{dyn}) is not of the Sturm-Liouville type.
\end{thebibliography}
\end{document}